# Towards an Ethical Framework in the Complex Digital Era

**David Pastor-Escuredo**[1,2,3,*] **and Ricardo Vinuesa**[4,5]

[1]*LifeD Lab, Madrid, Spain*

[2]*UCL, London, UK*

[3]*UNICEF*

[4]*FLOW Engineering Mechanics, KTH Royal Institute of Technology, Stockholm, Sweden*

[5]*KTH Climate Action Centre, Stockholm, Sweden*

[*]*corresponding author: david@lifedlab.org*

**Abstract.** The digital revolution has brought ethical crossroads of technology, behavior and truth. However, the need of a comprehensive and constructive ethical framework is emerging as digital platforms have been used to build a global chaotic and truth-agnostic system. The unequal structure of the global system leads to dynamic changes and systemic problems, which have a more significant impact on those that are most vulnerable. Ethical frameworks based only on the individual level are no longer sufficient as they lack the necessary articulation to provide solutions to the new challenges. A new ethical vision must comprise the understanding of the scales and complex interconnections, as well as the causal chains of modern social systems. Many of these systems are internally fragile and very sensitive to external factors and threats, which lead to unethical situations that require systemic solutions still centered on individuals. Furthermore, the multi-layered net-like social tissue generates clusters of power that prevent certain communities from proper development. Digital technology has also impacted at the individual level posing the risk of a more homogeneous, predictable and ultimately controllable humankind. To preserve the core of humanity and the aspiration of common truth, a new ethical framework must empower individuals and uniqueness, as well as cultural heterogeneity, tackling the negative outcomes of digitalization. Only combining human-centered and collectiveness-oriented digital development will it be possible to construct new social models and interactions that are ethical. This vision requires science to enhance ethical frameworks and principles using computational tools to support truth-grounded actions, so as to transform and configure properties of the social systems.

## 1 A historical overview

Ethics has always been a fundamental matter for humanity. From the first cultures where ethics was guided by the behavior of gods to the contemporary atheist cultures (Kramer, 1981; Masson-Oursel, 1938; OPPENHEIM, 1968). Evolving from the myth and the great chronical literature, Greek philosophers brought a new political and philosophical dimension to ethics (Hackforth, 1972) (Ameriks & Clarke, 2000). Throughout the Classical Culture of Greece and Rome, ethics was related to the cosmical and urban order. The medieval age was an enforced turn to religions all around the globe, and ethics was related to and constrained by religion.

The Enlightenment brought a new vision of ethics centered in humans and their reason and science with important contributions on idealism, Kant and Hegel transcendental philosophies and the beginning of Aesthetics as a discipline in Philosophy (Kant, 2002). Modernity started with the suspicion of a single truth which could be achieved by religion or human reason, specially Freud, Marx and Nietzsche (Freud, 2020; Marx & Engels, 2004; Nietsche, 1989). Modernity was the end of the big systems of philosophy because it was impossible to build a system based on human reason that could be justified through reason itself. The concept of life took more importance because of the Hermeneutical philosophers and the philosophers of Phenomenology, specially in Dilthey. Modernity is, at the end, the end of ethics and the opening of subjectivism. Post-modernity, was therefore characterized by a bunch of trends in Philosophy. The trust in reason lead to ethics that trusted in scientific truth (Positivism) or the truth that could be achieved through democratic systems ("Escuela de Frankfurt"). On the other hand, the vision of ethics without truth led to the different authors of Estructuralism with a very strong sense of subjectivism and also the authors that relied on Anthropology to observe different cultures as the basis for societal order. Finally, big philosophical systems found different authors that revisited the systems and proposed new innovations that tried lead the economics of the globe (Materialism). In this landscape, capitalism and economic globalization led to a mosaic of cultures around the globe with a strong sense of individuality (Sartre, Camus, Sprintzen, & Van den Hoven, 2004). As a result, ethics became highly fragmented and enclosed within communities of practice. Ethic was forced to establish their own limits (Muguerza, 2004). Lately, computational social science has gained traction by using big data to analyze human behaviors with practically no sense of ethics, only through the guidelines of United Nations for sustainable and fair societies.

Fortunately, through the last decades, technological development has become one of the main drivers of socio-economic configuration of the world map. The Internet enabled hyper-connectivity and real-time flows of information with strong implications in communication and financial systems (Castells, 2002). The awareness of the power of the data generated by the society brought the hope of a digital revolution (Kitchin, 2014). Also, Artificial Intelligence (AI) is reaching its expected potential (Simon, 1969) (Stone et al., 2016) having been the core of an industrial revolution (Schwab, 2017) which poses questions about the future path of humankind. An important issue is that, so far, technological development has gone in a faster pace than the evolution of ethic. The fragmented ethical landscape of AI guidelines is a significant problem to establish a truth-grounded governance of technology and its applications (Jobin, Ienca, & Vayena, 2019). Still, the guidelines by the United Nations, i.e. the Millenium Goals and the Sustainable Development Goals, are the most comprehensive sets of knowledge and rules to structure societal order.

The exponentially increasing need of addressing more specific and heterogeneous challenges in different parts of the world has led to the global commitment of the Sustainable Development Goals that (Sachs, 2012) (Griggs et al., 2013). Poverty, gender, inequality, vulnerability and how we live in our planet have become central issues for humankind and governments worldwide. The long-term development goals are also challenged by crisis of different kind and large-scale transformations that produce systemic changes. The dynamics of societies and the relationships with the planet become part of a framework that has strong ethical implications. In this scenario, the big-data revolution has been identified as the opportunity to drive all societal changes for a better world (Kirkpatrick, 2013). These challenges to tackle are a new opportunity to revitalize ethic in a more universal sense so that we succeed in building a better humanity (Apel, 2008). Integrating the data revolution and the area of Philosophy is definitely an opportunity to build new societal structures in an era of complexity and technology (C. F.-A. David Pastor-Escuredo, Jesus Salgado, Leticia Izquierdo y Mª de los Angeles Huerta, 2021; G. G. David Pastor-Escuredo, Julio Lumbreras, Juan Garbajosa, 2021; Pastor-Escuredo, 2020; Pastor-Escuredo & Tarazona, 2020).

## 2 A global challenge needs a systemic approach

Computational decisions and Data are the new elements of an emergent revolution, but this revolution goes beyond industrial. In a global scale, the interconnections between different global subsystems, the real-time information and the social response to information (and misinformation), events and crises make the world a tangled network of complex processes. At the individual level, the use of digital platforms, apps and tools facilitate peoples' lives in many aspects and generate massive amounts of behavioral data with very high-value and large ubiquity, provided artificial-intelligence (AI) tools to analyze it (UNGP; UNICEF). Thus, the local and the global scales are more interconnected than ever through Data. However, the topology of these interconnections is very clustered, unevenly layered and hierarchical, which translates into inequalities, power concentration, marginalization, vulnerability and fragility (Hilbert, 2016). An integrative and high-impact transformative action starts from making ethical principles become a reality.

For this transformation, a top-down process based on contrasted knowledge of human chains has to be integrated with a bottom-up process of inter-individual consensus. Evidence-based consensus can articulate a framework that applies to communities and discover commonalities and invariants within the heterogeneities to identify what principles should be universal and promoted (Jobin et al., 2019). Digitalization is then a subject of ethics as an induction tool, and computation can be a part of evidence-based ethical frameworks as a deduction tool.

To overcome the problems of industrialization and automation, ethical frameworks must embody conceptual and systemic risks. It is not only important to ensure the rights, but also the narratives that outlast and shape the perception and lives (Varela, Thompson, & Rosch, 2016). Digital technologies have proved to be a differential tool to create the mechanisms that protect rights and allow a faster recovery necessary to sustain development (Pulse, 2015, 2016; UNGP). However, the digitalization as a global process also poses important risks in terms of individual rights and values and cultural heterogeneity that cannot be neglected by design if we want to avoid industrialization inequalities.

Complexity, climate evolution and computational methods give rise to new systemic risks that require a construct to drive inter-individual consensus built from communities, starting from the so-called Collective Intelligence (Malone, 2004; Malone, Laubacher, & Dellarocas, 2009, 2010). In contrast with the current evolution of digitalization, ethics must guide how to build models of society according to a vision of humanity as the lens to define principles through innovation and deduction methodologies (Escuredo, Fernández-Aller, Salgado, Izquierdo, & Huerta, 2021). For instance, digitally-interconnected communities can use technologies to reach agreements and share them with other communities such as virtual congress (Chile, 2020). Reducing the risks associated with digitalization and augmenting social systems through digital technologies are complementary processes that must be undertaken under the light of ethics formulated not only as protection rights but also as actionable universal principles (Escuredo et al., 2021). These principles would not be meant to increase globalization of digitalization, but to promote actions at the local level to implement the necessary mechanisms according to the cultures. This is essential in order to ensure a more prosperous future enabled by novel digital technologies, while not compromising the achievement of SDG-10 on reduced inequalities (Vinuesa et al., 2020).

Cities is the SDG-11 of United Nations to make them more sustainable but also more inclusive and resilient. This SDG comprises 10 targets including small scale targets based on indicators and large financial and economic targets that evolve considering large-scale complexity.

## 3 Features of an ethical framework

In order to achieve the goal described above, ethics has to adopt concepts and language from science and develop shared knowledge and terminology in order to organically integrate the wisdom of tradition into the paradigm of global digitalization. Ethical principles such as truth, liberty or solidarity must be revisited to be inserted into digitalization at the very core of its design, development and deployment in the real world considering the current use of digital technologies.

Pro-active work to achieve the SDG-11 targets demands an ethical framework starting from principles, protective, actionable and projection principles (Escuredo et al., 2021). Protective principles are cybersecurity, anti-virality, integrity, privacy, legality, explicability, trustfulness, transparency-accessibility, accountability and no-discrimination. Actionable principles are democratization, impact-based, self-sustainability, literacy, digital inclusion, participation, capacity-building, digital solidarity-philanthropy, collaboration, robustness, anti-discrimination and responsibility. Projection principles are multiculturality, multi-level society, internationality, autonomous society, sustainability, resilience and sensibility-sensitivity. Having these principles coming to force as a framework it is possible to build soft-regulation and social tissues to make the SDG-11 targets turn true.

On the contrary, digitalization is being led by technical and industrial principles (Union, 2020)´, and a number of fragmented principles of conduct which can be denoted as practice guidelines. Data and AI are being used to make better decisions, optimize processes, customize services, predict patterns and understand society. As ethical questions emerge, several guidelines have been proposed to ease the ongoing AI penetration and also analyze the risks. In (Jobin et al., 2019), five principles were identified as a convergent set across multiple frameworks: transparency, justice and fairness, non-maleficence, responsibility and privacy.

In particular, transparency of AI models (Rudin, 2019) was identified as a key aspect that needs to be achieved with interpretability methods, as discussed by (Vinuesa & Sirmacek, 2021). However, universal ethical principles within transcendental ethics (Kant, 2002) must become central for a sustainable industry and digitalization as enumerated above. The misuse or missed use of digital technologies depend on understanding the catalyzing or inhibiting roles of technology (Vinuesa et al., 2020), so an ethical framework based on principles for new governance is necessary (Pastor-Escuredo & Treleaven, 2021)

Although there is an evident trend for ethical principles in digitalization and AI development, an ethical approach to a more comprehensive vision of humanity and socio-technological models is needed. To this end, some approaches have been proposed, such as Human Rights-based digitalization or the systematic integration of heterogeneous ethical frameworks (Jobin et al., 2019). These approaches rely on consensus, although they do not guarantee that the actual actions, projects and initiatives lead a more sustainable and better world. However, a global human vision is needed to make sense of the consensus. The Sustainable Development Goals (SDGs) is a step forwards on having such a vision, but the lack of a longer-term vision of humanity and the non-ethically-driven response to crises hamper the actual implementation of the SDGs and structures for building up the future and ethics-based governance (Pastor-Escuredo & Treleaven, 2021)

As SDG-11, all SDGs include targets affecting individuals (no hunger, no poverty, gender equality), but imply a direct systemic change in economies, relationships with the planet, social behavior and regulation. The relationship between digitalization and the SDGs is complex and bidirectional. Several projects in the field of sustainable development have shown the potential of digital innovation (Pulse, 2016; UNGP; UNICEF). Further innovation and deduction work based on AI and Data will lead to soft regulation and new governance systems based on the principles mentioned.

Ethical principles are the basis to articulate a comprehensive framework based on the SDGs. For this mission, two main processes have to be carried out. First, understanding the ethical implications of the actions, roadmaps and digital innovations needed to achieve the SDGs. Second, using and deducing structures from the principles that comprises protection, actionable principles and project the future as constitutive elements in policies and governance. Overcoming vulnerability and individual-level problems of the current world requires not only formal structures based on technology, but also use technology to shed light and truth as the basis for responsible governance. Considering this, beyond protection-oriented ethical principles that address misuse, actionable and future-projection principles have been proposed.

Revisiting Kant, his ethics based on the "categorical imperative" could be translated into protective and actionable principles. However, humanity has gone further since the Enlightment with principles such as sustainability and resilience that are projective and included in the SDGs. Thus, SDGs are a proper framework to propose ethical principles because of the targets that go from individual indicator-based targets to more complex targets that are conceived at larger scales. By deducing and innovating on SDGs and their targets and indicators through computational approaches it is possible to reach several scales and make ethical principles come into force and become a reality. Moreover, computational approaches take into consideration local (cultural) data, so the integration between universal principles and cultural differences is possible thanks to science-powered ethics and computation.

From a constructive point of view, computational approaches can be articulate ethical frameworks to draw conclusions that could not be made before due to the lack of evidence and large-scale inference processes. Computational intelligence can help remove biases, personal interests of power and make more transparent and fair transactions (G. G. David Pastor-Escuredo, Julio Lumbreras, Juan Garbajosa, 2021). This type of intelligence must be based on a hybrid approach involving both humans and machines, this means Collective Intelligence (Pastor-Escuredo & Treleaven, 2021). However, we lack the technology-governance entities and mechanisms to evaluate digitalization. There is a latent high risk in simplifying the science behind AI in the context of an industrialization process that impacts our society and well-being.

An ethical framework should address the configuration of the social tissue, soft regulation and new governance and how digital technologies are leveraged for public purposes or high-impact private platforms. The capacities to have these mechanisms and criticism are yet to be developed. Proactive actions based on collaboration, educational programs engaging practitioners and citizens, as well as Collective Intelligence initiatives and platforms are key to harness the visions of the future society, specially in city science and computational socio-economics and dynamics of the city (Escuredo et al., 2021; Pastor-Escuredo & Frias-Martinez, 2020).

## 4 Outlook

The quickly-growing development of machines cannot be longer motivated by interests to just develop smarter methods to perform human-like tasks. Traditionally, this vision has been denominated physicalism. Machines will hardly reproduce human reasoning, but in the meantime, they can be used in applications that de-humanize the society by a non-regulated industrialization that leads to a digital gap. Even tasks that are performed with accuracy such as text analysis, translation and generation, or even more creative tasks such as painting (LeCun, Bengio, & Hinton, 2015) still lack human characteristics such open interpretability and liberty.

On the other hand, humans are not perfect and machines can help reshape how power relationships lead to vulnerability and fragility of social systems. Computational frameworks are not only a way for efficiency but also transparency and accountability to implement better governance systems without human-centric biases. Even more, computations can help create different scenarios to improve decision making. This type of Collective Intelligence systems has to be built with ethics by design and also a more comprehensive human-machine relationship vision. Collective Intelligence presents several advantages. First of all, make humans invest more time in high-level and abstract thinking. Second, it powers-up machine workflows by introducing human skills and cognition. Third, it makes humans' life easier without a sense of industrialization. Finally, it allows to scale-up governance by integrating many people into machine workflows (Pastor-Escuredo & Treleaven, 2021).

We can expect that cultures will adopt and develop different levels and applications of computation because of their tradition, religion, industry and research system. We should start wondering about the social status of machines and how they will collaborate with humans to potentially take humanity to a next level of self-development and well-being. In this context, digital divides, and uneven access to AI resources, are key aspects increasing inequalities and worldwide, and therefore need to be accounted for when developing ethics-based platforms.

A new universal ethical framework that learns from the evidence that science provides is required. This may be one of the most powerful contributions that digital technologies can provide to the society. It is not sufficient that ethics reflects about science as a subject: science has to be part of ethics because it can help it to understand this complex world. One critical aspect of this mission in the short term is to understand the relationships between the different social scales and the SDGs that drive the sustainable development progress in the next few years. As an alternative to negative techno-ethics that are limited to protection principles, it is necessary to promote an ethical framework that creates a conceptual construct and vision capable to drive technological development towards a better humanity. This means complementing protection guidelines with actionable principles, and also with projection principles based on imagination, creativity and collective intelligence.

This landscape prompts for an exciting return to formulating questions about human conceptions and expectations by ethical principles.

### Acknowledgments

D.P.-E. would like to thank the Innovation and Technology for Development Centre (itdUPM). R.V. acknowledges the support by the Swedish Research Council (VR).